\documentstyle{article}
\catcode`\@=11

   \renewcommand{\section}%
   {\setcounter{equation}{0}\@startsection {section}{1}{\z@}{-3.5ex plus -1ex
   minus -.2ex}{2.3ex plus .2ex}{\Large\bf}}

   \newcommand{\beq}{\begin{equation}}
   \newcommand{\eeq}{\end{equation}}
   \newcommand{\beqs}{\arraycolsep1.5pt\begin{eqnarray}}
   \newcommand{\eeqs}{\arraycolsep5pt\end{eqnarray}}
   \newcommand{\beqsn}{\arraycolsep1.5pt\begin{eqnarray*}}
   \newcommand{\eeqsn}{\end{eqnarray*}\arraycolsep5pt}
   \newcommand{\bmatrix}{\arraycolsep5pt\begin{array}}
   \newcommand{\ematrix}{\arraycolsep1.5pt\end{array}}

 \def\bbbone{{\mathchoice {\rm 1\mskip-4mu l} {\rm 1\mskip-4mu l}
 {\rm 1\mskip-4.5mu l} {\rm 1\mskip-5mu l}}}

 \def\bbbr{{\rm I\mskip -3.5mu R}}

 \def\bbbone{{\mathchoice {\rm 1\mskip-4mu l} {\rm 1\mskip-4mu l}
 {\rm 1\mskip-4.5mu l} {\rm 1\mskip-5mu l}}}

 \def\bbbr{{\rm I\mskip -3.5mu R}}

\catcode`\@=12

\begin{document}
\title{ Soliton solutions of the Hamiltonian DSI and DSIII equations
\thanks{Work supported in part by M.U.R.S.T -- Italy}}

\author{Flora Pempinelli\\
Dipartimento di Fisica dell'Universit\`a e Sezione INFN\thanks{e-mail:
PEMPI@LECCE.INFN.IT}\\
73100 LECCE --  Italy}
\maketitle
\thispagestyle{empty}

\begin{abstract}
By introducing generalized B\"acklund Transformations depending on arbitrary
functions, wave and localized soliton solutions of the Davey-Stewartson
equations are generated. Moreover explicit soliton solutions of the
Hamiltonian DSI  and DSIII equations are obtained.
\end{abstract}

\section{Introduction}

Many efforts have been made in the last years to extend the soliton theory
to the non linear evolution equations (NLEEs) in 2 + 1 (two spatial
 and one temporal) dimensions. The Spectral Transform (ST), which is the
principal mathematical tool of the soliton theory, has been extended to
dispersive NLEEs in 2 + 1 dimensions in the first '80 years, but it was
generally admitted the lack of two dimensional localized solitons. Only
recently, in 1988, it has been discovered  by Boiti-L\'eon-Martina-
Pempinelli [\ref{BLMP88}] that all the
equations in the hierarchy related to the Zakharov-Shabat (ZS)
hyperbolic spectral problem in the plane have exponentially localized
soliton solutions. The most representative equation in the hierarchy is the
Davey-Stewartson I (DSI) equation, which provides a two dimensional
generalization of the non linear Schr\"odinger (NLS) equation.  These
2-dimensional soliton solutions (also called dromions [\ref{FS90}])
 display a richer
phenomenology than in 1 + 1 dimensions. The scattering of the solitons can be
inelastic and they can change shape and also exchange mass. They can also
simulate inelastic scattering processes of quantum particles as creation
and annihilation, fusion and fission, and interaction with virtual
particles [\ref{bifur},\ref{dynamics}]. Moreover, as a relevant application
of the spectral method proposed by Sabatier [\ref{Sabatier}] and further
developed by Boiti-Pempinelli-Sabatier [\ref{BPS}], it has been stated that
an additional NLEE, called Davey-Stewartson III (DSIII) equation
admits also localized soliton solutions with properties similar to those of
the DSI equation [\ref{Pempiexeter}]. Localized soliton solutions for the DSI
equation were first found by using gauge B\"acklund Transformations (BT)
[\ref{BLMP88},\ref{JMP}]. Later these solutions were also obtained by means of
the Inverse Spectral Transform (IST) [\ref{FS89},\ref{FS90},\ref{bifur},\ref
{dynamics}] and direct methods
 [\ref{HietHir},\ref{MartAlonso},\ref{GilsonNimmo}]. Now we want to reobtain
them by using a generalized gauge BT [\ref{waves}]. The advantage of this
method is that in such a way it is possible to obtain soliton solutions for
DSI and DSIII equations in the so-called Hamiltonian case. For this very
interesting case we have not presently at our disposal the ST or the
dressing method, so to get explicit solutions we must generalize the
B\"acklund gauge in order to include the special form of the boundaries in
the Hamiltonian case.

Finally let us note that a review on multidimensional localized solitons
and more references are given in [\ref{review}].

\section{Davey-Stewartson I and III equations}

Let us write the DSI equation in his more general two component form in
characteristic coordinates

 \beqs
\label{DSIQ}
&&iQ_t+\sigma_3(Q_{uu}+Q_{vv})+[A,Q]=0\\
\label{DSIA}
&&\left(\bmatrix{cc}A^{(1)}_u&0\\0&A^{(2)}_v\ematrix\right)=-{1\over2}
\sigma_3\left(\bmatrix{cc}{(Q^2)}_v&0\\0&{(Q^2)}_u\ematrix
\right)
\eeqs
where
\beq
\label{QA}
 Q=\left(\bmatrix{cc}0&q(u,v,t)\\r(u,v,t)&0\ematrix\right),\quad
   A=\left(\bmatrix{cc}A^{(1)}(u,v,t)&0\\0&A^{(2)}(u,v,t)\ematrix\right)
\eeq
$u$ and $v$ are the characteristic coordinates $u=x+y$, $v=x-y$.

The equation DSI is compatible with the reduction
\beq
\label{red}
q=\varepsilon \overline{r},\quad \varepsilon = \pm
\eeq
(where $\overline r$ denotes the complex conjugate of $r$), which furnishes
 the so-called reduced DSI equation, describing physical situations as in
hydrodynamics or in plasma physics.

The equation DSI can be obtained [\ref{BPS}, \ref{Pempiexeter}] as the
compatibility condition between two Lax operators $T_1$ and $T_2$ which
commute in the ``weak" sense [\ref{BLMannaP}]
\beqs
\label{weak}
T_1\varphi=0,\qquad[T_1,T_2]\varphi=0.
\eeqs
$T_1$ is the ZS hyperbolic operator in the plane
\beq
\label{ZS}
T_1\varphi\equiv\left\{2\left(\bmatrix{cc}\partial_u&0\\0&\partial_v
\ematrix\right)+Q\right\}\varphi=0
\eeq
and $T_2$ has the form
\beq
\label{DSIT0}
2T_2\varphi\equiv\left\{i\partial_t+\partial_v^2-\partial_u^2+A+
\left(\bmatrix{cc}0&-q_u\\r_v&0\ematrix\right)\right\}\varphi=
-k^2\varphi\sigma_3.
\eeq

Let us remember that the boundary conditions of the auxiliary field $A$ can
be arbitrarily chosen. The equation DSI in his standard version, i.e. with the
boundary written as
\beqs
A^{(1)}(u,v,t)&=&-{1\over2}\!\!\int^u_{-\infty}\!\!du'(Q^2)_v+a_0^{(1)}(v,t)\\
A^{(2)}(u,v,t)&=&{1\over2}\!\!\int^v_{-\infty}\!\!dv'(Q^2)_u+a_0^{(2)}(u,t)
,
\eeqs
admits localized soliton solutions [\ref{BLMP88}]. The one-soliton solution
has the form
\beqs
q=-\frac{2}{D}\lambda_I\eta e^{i\theta},\qquad r=
-{2\over D}\mu_I\rho e^{-i\theta}
\eeqs
where
\beqs
&&D=2\gamma(\cosh \xi_1+\cosh\xi_2)+e^{\xi_2}\\
&&\xi_1=-\mu_Iu-\lambda_Iv+2(\lambda_R\lambda_I+\mu_R\mu_I)t\\
&&\xi_2=\mu_Iu-\lambda_Iv+2(\lambda_R\lambda_I-\mu_R\mu_I)t\\
&&\theta=\mu_Ru+\lambda_Rv+(\lambda_I^2-\lambda^2_R+\mu_I^2-\mu_R^2)t\\
&&\gamma={1\over4}\eta\rho.
\eeqs
The complex parameters $\lambda=\lambda_R+i\lambda_I$, $\mu=\mu_R+i\mu_I$
are
the discrete eigenvalue of the associated Zakharov--Shabat spectral problem
and $\rho$, $\eta$ are arbitrary complex constants satisfying the
conditions
$\gamma\in\bbbr$ and $\gamma(1+\gamma)>0$.

Another relevant choice for the boundaries can be done
\beqs
\label{HamiltonADSI}
&&A^{(1)}=-{1\over4}\left(\int_{-\infty}^u+\int_{+\infty}^u\right)
du'\,(Q^2)_v+A_0^{(1)}(v,t)\\
&&A^{(2)}={1\over4}\left(\int_{-\infty}^v+\int_{+\infty}^v\right)
dv'\,(Q^2)_u+A_0^{(2)}(u,t).\nonumber
\eeqs
In this case one can introduce the Hamiltonian
\beq
H=\!\!\int\!\!\!\int\!\!dudv\left[r(\partial_u^2+\partial_v^2)q-{1\over4}qr
(\partial_u\partial_v^{-1}+
\partial_v\partial_u^{-1})qr+(A_0^{(1)}-A_0^{(2)})qr\right]
\eeq
and the canonical Poisson bracket
\beq
\{F,G\}=i\!\!\int\!\!\!\int\!\!dudv\left[
{\delta F\over\delta q}{\delta G\over\delta r}-
{\delta F\over\delta r}{\delta G\over\delta q}\right]
\eeq
where $q$ and $r$ are the conjugate variables. Then the equations of motion
\beq
q_t=\{q,H\},\qquad r_t=\{r,H\}
\eeq
yield the DSI equation. It has been shown that the DSI equation for
 $A_0^{(1)}\equiv A_0^{(2)}\equiv 0$ is completely integrable in the
hamiltonian
 sense [\ref{ShAbl}, \ref{KulLip}], so this case is called Hamiltonian.
Quantum extensions have also been
done. However the problem of defining a ST is completely open.

There is another NLEE that can  be associated to ZS hyperbolic spectral
operator in the plane and admits localized soliton solutions, the so-called
DSIII equation [\ref{BPS},\ref{Pempiexeter}]. Indeed the hamiltonian version
of this equation appears already in [\ref{Shul}] and a bihamiltonian version
in  [\ref{SF}]. The DSIII equation has the form
\beqs
&&iQ_t+\sigma_3(Q_{vv}-Q_{uu})+[A,Q]=0\\
\label{DSIIIA}
&&\left(\bmatrix{cc}A_u^{(1)}&0\\0&A_v^{(2)}\ematrix\right)=-{1\over2}
\left(\bmatrix{cc}{(Q^2)}_{v}&0\\0&{(Q^2)}_{u}\ematrix\right).
\eeqs
Also this equation is compatible with the reduction $q=\varepsilon\overline r$.
It can
be obtained as the ``weak" compatibility condition between two operator
$T_1$ and $T_2$. $T_1$ is given, as before, by (\ref{ZS}). $T_2$ in this
case takes the form
\beqs
\label{DSIIIT0}
2T_2\varphi\equiv\left\{i\partial_t+\partial_u^2+\partial_v^2+A+\left(
\bmatrix{cc}0&q_u\\r_v&0\ematrix\right)\right\}\varphi=-k^2\varphi.
\eeqs
Also the DSIII equation in his standard version, i.e. with the boundaries
chosen as
\beqs
A^{(1)}(u,v,t)&=&-{1\over2}\!\!\int^u_{-\infty}\!\!du'(Q^2)_v+a_0^{(1)}(v,t)\\
A^{(2)}(u,v,t)&=&-{1\over2}\!\!\int^v_{-\infty}\!\!dv'(Q^2)_u+a_0^{(2)}(u,t
),
\eeqs
admits localized soliton solutions [\ref{BPS}, \ref{Pempiexeter}], that
have the same shape of those of the DSI equation, but different time evolution.
The one-soliton solution has the form

\beqs
q=-\frac{2}{D}\lambda_I\eta e^{i\theta},\qquad r=
-{2\over D}\mu_I\rho e^{-i\theta}
\eeqs
where
\beqs
&&D=2\gamma(\cosh \xi_1+\cosh\xi_2)+e^{\xi_2}\\
&&\xi_1=-\mu_Iu-\lambda_Iv+2(\lambda_R\lambda_I-\mu_R\mu_I)t\\
&&\xi_2=\mu_Iu-\lambda_Iv+2(\lambda_R\lambda_I+\mu_R\mu_I)t\\
&&\theta=\mu_Ru+\lambda_Rv+(\lambda_I^2-\lambda^2_R-\mu_I^2+\mu_R^2)t\\
&&\gamma={1\over4}\eta\rho.
\eeqs
As in the DSI case, the complex parameters
$\lambda=\lambda_R+i\lambda_I$, $\mu=\mu_R+i\mu_I$ are
the discrete eigenvalue of the associated Zakharov--Shabat spectral problem
and $\rho$, $\eta$ are arbitrary complex constants satisfying the
conditions
$\gamma\in\bbbr$ and $\gamma(1+\gamma)>0$.

Moreover the choice for the boundaries
\beqs
\label{HamiltonADSIII}
&&A^{(1)}=-{1\over4}\left(\int_{-\infty}^u+\int_{+\infty}^u\right)du'\,
(Q^2)_v+A_0^{(1)}(v,t)\\
&&A^{(2)}=-{1\over4}\left(\int_{-\infty}^v+\int_{+\infty}^v\right) dv'\,
(Q^2)_u+A_0^{(2)}(u,t)\nonumber
\eeqs
furnishes the Hamiltonian case. If we introduce the Hamiltonian
\beq
H=\!\!\int\!\!\!\int\!\!dudv\left[r(-\partial_u^2+\partial_v^2)q+{1\over4}qr
(\partial_u\partial_v^{-1}-
\partial_v\partial_u^{-1})qr+(A_0^{(1)}-A_0^{(2)})qr\right],
\eeq
the equations of motion $q_t=\{q,H\}$ and $r_t=\{r,H\}$ yield the DSIII
equation.

Finally let us make some comments. The DSI and DSIII equations belong to the
same
principal spectral problem (\ref{ZS}), so the direct and the inverse
problems are the same for these two equations. On the contrary, they have
different auxiliary spectral problems, so the evolution of the ST and the
dispersion relation are different. Also, as we have seen, there is a simple
modification for the boundary conditions. The conclusion is that {\it all
the results} known for the DSI equation can be easily {\it extended} to the
DSIII
equation [\ref{Pempiexeter}].


\section{Solutions via B\"acklund Transformations}
Given a solution $Q$ of the DSI (or DSIII) equation, we want to generate a new
solution $Q'$ of the same equation by introducing a convenient gauge
operator $B$ which transforms by means of
\beq
\label{gaugeB}
\psi'=B(Q',Q)\psi
\eeq
the matrix solution $\psi$ of the principal spectral problem
\beq
T_1(Q)\psi=0
\eeq
for $Q$ to the matrix solution $\psi'$ of the same spectral problem for
$Q'$
\beq
T_1(Q')\psi'=0.
\eeq
It is easy to verify that if $B$ satisfies
\beqs
\label{spaceBacklund}
T_1(Q')B(Q',Q)-B(Q',Q)T_1(Q)=0\\
\label{timeBacklund}
T_2(Q')B(Q',Q)-B(Q',Q)T_2(Q)=0,
\eeqs
then $T_1(Q')$ and $T_2(Q')$ satisfy the same compatibility condition and
therefore $Q'$ satisfies the same equation as $Q$. The above equations
furnish, respectively, the so-called space and time component of the BT.

We are interested in the most general B\"acklund gauge as polynomial of the
first order in $\partial_y$
\beq
B(Q',Q)=\alpha\partial_y+B_0(Q',Q)
\eeq
with $\alpha$ a constant diagonal matrix and $B_0$ a matrix. By inserting
it in (\ref{spaceBacklund}), where $T_1$ is the ZS hyperbolic spectral
operator (\ref{ZS}), we get
\beq
\label{fullBacklund}
B(Q',Q)=\alpha\partial_y-\hbox{$1\over2$}\sigma_3(Q'\alpha-\alpha Q)
-\hbox{$1\over2$}\sigma_3\alpha\, {\cal I}(Q'^2-Q^2)+\beta
\eeq
and the space component of the B\"acklund transformation
\beqs
Q'\big[\beta-\hbox{$1\over2$}\alpha\sigma_3\,{\cal I}(Q'^2-Q^2)\big]
-\big[\beta-\hbox{$1\over2$}\alpha\sigma_3\,{\cal
I}(Q'^2-Q^2)\big]Q\nonumber\\
{ }-\hbox{$1\over2$}\sigma_3(Q'\alpha-\alpha Q)_x
-\hbox{$1\over2$}(Q'\alpha+\alpha Q)_y=0.\label{spacecomponent}
\eeqs
The matrix operator ${\cal I}$ is defined by
\beq
{\cal I}=(\partial_x+\sigma_3\partial_y)^{-1}
\eeq
and the diagonal matrix $\beta$ is subjected to the constraint
\beq
\label{beta}
(\partial_x+\sigma_3\partial_y)\beta=0,
\eeq
i.e. it is of the form
\beq
\beta=
\left(\matrix{\beta_1(v,t)  &       0        \cr
                    0         & \beta_2(u,t) \cr}  \right)
\eeq
where $\beta_1$ and $\beta_2$ are arbitrary functions. Note that {\it for
$\beta$ a space dependence is allowed} in contrast with the 1 + 1 dim case
where $\beta$ is the constant of integration. {\it This freedom will be
used for getting soliton solutions for Hamiltonian DSI and DSIII equations.}
By inserting (\ref{fullBacklund}) in (\ref{timeBacklund}) we get the time
component of the BT, which is equivalent to the DSI (or DSIII) equation for
$Q'$ plus two additional equations that can be used for determining the
auxiliary field $A'$ and the admissible $\beta$'s. For details for the DSI
equation see [\ref{waves},\ref{review}]. The final result is that $\alpha$
and $\beta$ can be written as
\beq
\alpha={\bbbone },
\qquad \beta=
\left(\matrix{\lambda(v,t)  &  0 \cr
                   0        &  \mu(u,t) \cr}
\right)
\eeq
in the B\"acklund gauge $B=B(Q,Q';\lambda,\mu)$. Note that $\lambda$ and
$\mu$ generalize the parameters of the BT in 1 + 1 dimensional case.

For $\lambda$ and $\mu$ complex constants, the BT furnishes the localized
soliton solutions for DSI (or DSIII) equation for the choice $Q=0$, $A=0$
[\ref{BLMP88}, \ref{JMP}].

But to obtain soliton solutions for the Hamiltonian DSI (or DSIII) equation,
 we must use
\begin{enumerate}
\item the space dependence of $\lambda$ and $\mu$
\item $A$ can be $\neq 0$ when $Q=0$.
\end{enumerate}

In the case where $\lambda$ and $\mu$ have a space dependence, the
calculations are very complicated and some constraint equations appear. We
will give here only the principal results. Details for the DSI case can
be found in
[\ref{waves}]. We will restrict here at this equation. Results for the
DSIII equation can be obtained with some modifications from those of the DSI
equation.

The starting solution is $Q=0$ and $A=\mbox{diag}(A_{00}^{(1)}(v,t),
 A_{00}^{(2)}(u,t))$ where $A_{00}^{(i)}$ are arbitrary boundary values. If we
choose $A_{00}^{(i)}$ real and moving with constant speed
\beqs
&&A_{00}^{(1)}(v,t)=A_{00}^{(1)}(v+2\phi t) \nonumber\\
&&A_{00}^{(2)}(u,t)=A_{00}^{(2)}(u+2\theta t),
\label{3.62}
\eeqs
a new solution $Q$, $A$ can be explicitly written
\beq
q=\frac{W(\eta,{\cal D})}{{\cal
ED}+\rho\eta/4}\exp[-i(\epsilon+\delta)],\qquad
r=-\frac{W(\rho,{\cal E})}{{\cal
ED}+\rho\eta/4}\exp[i(\epsilon+\delta)]                \label{3.68}
\eeq
\beq
\bmatrix{c}
A^{(1)}=A_{00}^{(1)}+
2\partial_v^2\log({\cal ED}+\rho\eta/4),\\
A^{(2)}=A_{00}^{(2)}-
2\partial_u^2\log({\cal ED}+\rho\eta/4)
\ematrix                                             \label{3.69}
\eeq
with  boundary values
\beq
\bmatrix{c}
A_0^{(1)}=A_{00}^{(1)}+
\partial_v^2\log({\cal ED}+\rho\eta/4)\vert_{u=-\infty}+
\partial_v^2\log({\cal ED}+\rho\eta/4)\vert_{u=+\infty},\\
A_0^{(2)}=A_{00}^{(2)}-
\partial_u^2\log({\cal ED}+\rho\eta/4)\vert_{v=-\infty}-
\partial_u^2\log({\cal ED}+\rho\eta/4)\vert_{v=+\infty}
\ematrix                                             \label{3.70}
\eeq
and
\beq
\delta=\phi v+(\phi^2-\phi_0)t+\delta_0,\quad
\epsilon=\theta u+(\theta^2-\theta_0)t+\epsilon_0,
\eeq
where $\phi_0$, $\theta_0$, $\delta_0$, $\epsilon_0$ are real constants.
For the reduction $r=\varepsilon \overline q$, the wronskian $W$ becomes
constant
\beq
W(\eta,{\cal D})=2a, \quad W(\rho,{\cal E})= - 2 \varepsilon a,\qquad a\in\bbbr
{}.
\eeq
The real functions $\cal D$, $\eta$ and $\cal E$, $\rho$ satisfy the stationary
Schr\"odinger equations
\beq
\bmatrix{c}
{\cal D}_{vv}+(A_{00}^{(1)}-\phi_0){\cal D}=0,\\
\eta_{vv}+(A_{00}^{(1)}-\phi_0)\eta=0
\ematrix                                             \label{07}
\eeq
and
\beq
\bmatrix{c}
{\cal E}_{uu}-(A_{00}^{(2)}+\theta_0){\cal E}=0,\\
\rho_{uu}-(A_{00}^{(2)}+\theta_0)\rho=0.
\ematrix                                             \label{08}
\eeq
The localized soliton solution is reobtained with a special choice of the
functions $\cal D$, $\eta$, $\cal E$, $\rho$ and with $A_{00}^{(1)}=
A_{00}^{(2)}=0$. We are interested in the {\it Hamiltonian case}
\beq
A_0^{(1)}\equiv A_0^{(2)}\equiv 0
\eeq
in the reduced case. Let us introduce two functions
$\eta_0(v+2\phi t)$
and $\rho_0(u+2\theta t)$
 defined as follows
\beq
A_{00}^{(1)}=2\partial_v^2\log\eta_0,\qquad
A_{00}^{(2)}=-2\partial_u^2\log\rho_0.
\label{04}
\eeq
We have to solve a complicated nonlinear system of coupled equations with
constraints for $\cal D$, $\eta$, $\eta_0$ and $\cal E$, $\rho$, $\rho_0$.
If we add the additional constraints
\beqs
&&\lim_{u\to\pm\infty}\frac{\rho}{\cal E}=
-2(\rho_1\pm\rho_2),\label{011}\\
&&\lim_{v\to\pm\infty}\frac{\eta}{\cal D}=
-2(\eta_1\pm\eta_2)
\label{012}
\eeqs
($\rho_i$ and $\eta_i$ are real constants to be determined), the system
decouples and we obtain the solution
\beq
q=\frac {2a\rho_2\eta_2\rho_0\eta_0\exp[-i(\epsilon+\delta)]}
{\sinh\alpha\sinh\beta+(\rho_1\sinh\alpha+\sigma'\rho_2\cosh\alpha)
(\eta_1\sinh\beta+\sigma''\eta_2\cosh\beta)}
\label{026}
\eeq
where $(\sigma')^2=1$, $(\sigma'')^2=1$ and $\alpha$, $\beta$ are related
to $\cal D$ and $\cal E$ respectively. $\eta_0$, $\alpha$, $\rho_0$,
$\beta$ are determined by
\beqs
\bmatrix{ll}
\partial_v^2\eta_0+a^2\rho_2^2\eta_0^5-\phi_0\eta_0=0, &
 \partial_v\alpha=\sigma'a\rho_2\eta_0^2.\\
\partial_u^2\rho_0+a^2\eta_2^2\rho_0^5-\theta_0\rho_0=0, &
\partial_u\beta=-\sigma_0\sigma''a\eta_2\rho_0^2
\ematrix
\eeqs
with the consistency conditions
\beqs
&&\lim_{v\to\pm\infty}(\rho_1+\sigma'\rho_2
\coth\alpha)=-(\eta_1\pm\eta_2)^{-1},\label{024}\\
&&\lim_{u\to\pm\infty}(\eta_1+\sigma''\eta_2
\coth\beta)=-(\rho_1\pm\rho_2)^{-1}.
\label{025}
\eeqs
The  above ordinary differential equations for $\eta_0$ and $\rho_0$ can be
explicitly integrated in terms of elementary or classical trascendental
functions and it is easy to verify the consistency conditions.

Let us consider two cases in particular
\begin{enumerate}
\item \beq
\partial_v\eta_0\equiv0,\quad \partial_u\rho_0\equiv0, \quad
\phi_0=\lambda^2_0>0, \quad \theta_0=\mu^2_0>0.
\eeq

We get
\beq
q(u,v,t)=2|\lambda_0\mu_0|^{1/2}\frac
{\exp[-i(\epsilon+\delta)]}{\cosh\xi}
\label{029}
\eeq
where
\beqs
&&\xi=\mu_0(u+2\theta t-u_0)-\lambda_0(v+2\phi t-v_0),\nonumber\\
&&\epsilon=\theta u+(\theta^2-\mu_0^2)t+\epsilon_0,\nonumber\\
&&\delta=\phi v+(\phi^2-\lambda_0^2)t+\delta_0,
\label{030}
\eeqs
so $q$ in this case is a wave soliton solution.
\item \beq
\partial_v\eta_0\not\equiv0,\qquad
\partial_u\rho_0\not\equiv0,\qquad
\phi_0<0,\qquad \theta_0<0.
\eeq
In this case the solution can be explicitly written [\ref{waves}] in terms
 of the $\wp$,
$\zeta$, $\sigma$-Weierstrass elliptic functions and describes an infinite
wave with a periodically modulated amplitude.
\end{enumerate}


\subsection*{references}
\begin{enumerate}

\item
\label{BLMP88}
Boiti M, L\'eon J, Martina L and Pempinelli F 1988 {\it Phys. Lett.\/}
{\bf A 132} 432

\item
\label{FS90}
Fokas A S, Santini P M 1990 {\it Physica\/} {\bf D 44} 99

\item
\label{bifur}
Boiti M, L\'eon J, Pempinelli F 1990 {\it Inverse Problems\/} {\bf 6} 715

\item
\label{dynamics}
Boiti M, Martina L, Pashaev O K and Pempinelli F 1991 {\it Phys. Lett.\/}
{\bf A 160} 55

\item
\label{Sabatier}
Sabatier P C 1992 {\it Inverse Problems\/} {\bf 8} 263; Sabatier P C
1992 {\it Phys. Lett.\/} {\bf A 161} 345

\item
\label{BPS}
Boiti M, Pempinelli F, Sabatier P C 1993 {\it Inverse Problems\/} {\bf 9} 1

\item
\label{Pempiexeter}
Pempinelli F 1993 Localized soliton solutions for Davey-Stewartson I and
Davey-Stewartson III equations. In {\it Application of analytic and
geometric methods to nonlinear differential equations\/}, Clarkson P A
ed., pp 207-215, NATO ASI Series C 413, Kluwer Acad. Publ., Dordrecth

\item
\label{JMP}
Boiti M, L\'eon J, Pempinelli F 1990 {\it Jour. Math. Phys.\/} {\bf 31}
2612

\item
\label{FS89}
Fokas A S, Santini P M 1989 {\it Phys. Rev. Lett.\/} {\bf 63} 1329

\item
\label{HietHir}
Hietarinta J, Hirota R 1990 {\it Phys. Lett.\/} {\bf A 145} 237

\item
\label{MartAlonso}
Hernandez Heredero R, Martinez Alonso L, Medina Reus E 1991 {\it Phys.
Lett.} {\bf A 152} 37

\item
\label{GilsonNimmo}
Gilson C R, Nimmo J J C 1991 {\it Proc. R. Soc. Lond\/} {\bf A 435} 339

\item
\label{waves}
Boiti M, L\'eon J, Pempinelli F 1991 {\it Inverse Problems\/} {\bf 7} 175

\item
\label{review}
Boiti M, Martina M, Pempinelli F 1993 Multidimensional Localized Solitons,
to appear in {\it Chaos, Solitons and Fractals\/}

\item
\label{BLMannaP}
Boiti M, L\'eon J, Manna M and Pempinelli F 1986 {\it Inverse Problems\/}
{\bf 2} 271; Boiti M, L\'eon J, Manna M and Pempinelli F 1987 {\it Inverse
Problems\/} {\bf 3} 25

\item
\label{ShAbl}
Schultz C L, Ablowitz M J 1987 {\it Phys. Rev. Lett.\/} {\bf 59} 2825

\item
\label{KulLip}
Kulish P P, Lipovsky V D 1988 {\it Phys.Lett.\/} {\bf A 127} 413

\item
\label{Shul}
Shul'man E I 1984 {\it Theor. Math. Phys.} {\bf 56} 720

\item
\label{SF}
Santini P M, Fokas A S 1988 {\it Comm. Math. Phys.\/} {\bf115} 375

\end{enumerate}

   \end{document}